\documentclass[12pt]{article}
\usepackage{graphics}
\input epsf

\newcommand{\be}{\begin{eqnarray}}
\newcommand{\ee}{\end{eqnarray}}

\begin{document}
\title{Quarkonium Polarization in Lepton Annihilation}
\author{L. Clavelli\footnote{lclavell@bama.ua.edu}, 
P. Coulter\footnote{pcoulter@bama.ua.edu}, 
and
T. Gajdosik\footnote{garfield@bama.ua.edu},\\ 
Department of Physics and Astronomy, University of Alabama,\\
Tuscaloosa AL 35487\\}
\maketitle

\begin{abstract}
We investigate the polarization properties 
of vector quarkonia produced in lepton anti-lepton annihilation
with special attention to effects due to intermediate Z 
bosons.  
\end{abstract}

\renewcommand{\theequation}{\thesection.\arabic{equation}}
\renewcommand{\thesection}{\arabic{section}}

\section{\bf Introduction}
\setcounter{equation}{0}
    With the possible availability of $J/\Psi$ factories in the
foreseeable future, it is interesting to consider the discovery
potential of such machines for new physics effects
or, equivalently, for precision tests of the standard model \cite{weak}.
In this article we calculate the polarisation of $J/\Psi$ and 
other vector quarkonia in
resonant production through lepton-antilepton annihilation.
If the production proceeds through a single photon, the vector
quarkonium is an equal mixture of left and right handed polarization.
We also consider the production contribution through a virtual $Z$
leading to calculable deviations from this polarization.
As a preliminary application we derive the forward backward asymmetry
of the quarkonium decay into a lepton pair.  In section II we derive
the probabilities that the quarkonium state has each of the
three possible polarizations.  In section III we consider the forward
backward asymmetries.

     In its rest frame a massive vector meson has three possible
polarization states along any quantization axis.
\be
   \epsilon(\Uparrow)_\mu &=& -(0,1,+i,0)/{\sqrt {2}}
\label{Up}
\ee
\be
   \epsilon(\Downarrow)_\mu &=& (0,1,-i,0)/{\sqrt {2}} 
\label{Down}
\ee
\be
   \epsilon(0)_\mu &=& (0,0,0,1)
\label{Zero}
\ee

    The massive vector field has the canonical form

\be
  \Phi(x)_\mu = \frac{1}{{\sqrt{2E}} (2 \pi)^{3/2}}{\sum_{P,\lambda}}\left( \epsilon_\mu{(P,\lambda)} a(P,\lambda) e^{-iP \cdot x} 
+ \epsilon_{\mu}^ *(P,\lambda) a^\dagger(P,\lambda) e^{iP \cdot x} \right)
\ee

    In the zero binding approximation, the production of a quark with momentum
$P/2$ and spin projection $\lambda/2$ and an antiquark with the same momentum
$P/2$ and spin projection ${\overline \lambda/2}$ with $P^2 = M^2$, the
 vector meson mass squared, is proportional to the production of the vector 
 quarkonium with spin projection $\lambda/2 + {\overline \lambda/2}$.
Beginning in 1975 \cite{CSM}, various production and decay amplitudes have been 
discussed in this color singlet, zero-binding model. 
In this approximation, the quark mass and momentum are taken to be one half of 
the corresponding quarkonium
mass and momentum.
 Quantitatively, we have the identities \cite{qbarq}

\be
    v_\alpha(P/2, \uparrow) {\overline u}_\beta(P/2, \uparrow) &=& 
      \frac{1}{2 { \sqrt 2}} \left( \not\!\epsilon^*(\Uparrow) (\not\! P + M) \right)
      _{\alpha \beta}
\ee
\be
    v_\alpha(P/2, \downarrow) {\overline u}_\beta(P/2, \downarrow) &=& 
      \frac{1}{2 {\sqrt 2}} \left( \not\!\epsilon^*(\Downarrow) (\not\! P + M) \right)_{\alpha \beta}
\ee
\be
    \frac{1}{\sqrt{2}} { \left(
    v_\alpha(P/2, \uparrow) {\overline u}_\beta(P/2, \downarrow) 
    +v_\alpha(P/2, \downarrow) {\overline u}_\beta(P/2, \uparrow) \right)} 
    =
      \frac{1}{2 {\sqrt {2}}} \left( \not\! \epsilon^*(0) (\not\! P + M) \right)_{\alpha \beta}
.
\label{spinidentities}
\ee
These are readily derivable in the quark and antiquark rest frame which, in the zero
binding approximation, is the same as the quarkonium rest frame.  Then covariance leads
us to the above equations in any frame. We use the Bjorken-Drell conventions for
spinors, gamma matrices and Lorentz metric except that spinors are normalized to
${\overline u}u=2m$.
 In particular, $\epsilon_{0123}=1$.

     The color singlet model is then defined by inserting on the quark line
a wave function factor
\be
      F_W = \frac{T^0 \Psi(0)^*}{\sqrt{12 M}}\not\!\epsilon^* (\not\!P + M)
\ee

when a vector quarkonium is created and the conjugate wave function factor

\be
     \overline{ F}_W = \frac{\Psi(0) T^0}{\sqrt{12 M}}(\not\!P + M)\not\!\epsilon 
\label{wffactor}
\ee
when the vector quarkonium state decays.
Here $T^0$ is the $3 \times 3$ unit matrix in color space and $\Psi(0)$ is the
wave function averaged over a small volume around the origin.  This can be estimated
from the leptonic decay of the $J/\Psi$.
\be
     |\Psi(0)|^2 = \Gamma(J/\Psi \rightarrow e^+ e^-)M^2/(16 \pi \alpha^2 Q_q^2) \approx
  .0044 GeV M^2
\ee
The same wave function at the origin squared, scaling as $M^2$, also adequately describes
the $\Phi$ and $\Upsilon$ leptonic decay.

\section{\bf Vector Quarkonium Polarization in Lepton-Antilepton Annihilation}
\setcounter{equation}{0}

    In the zero-binding, color singlet model the polarization state of a produced 
vector quarkonium
is defined by the probabilities to produce at the resonant energy with no
relative momentum the corresponding combined spin state of
the quark and antiquark.  These probabilities are in turn proportional to the
squared production amplitudes.
In lepton-antilepton annihilation the production amplitudes are
\be
    {\cal M}(\lambda,{\overline \lambda}) &=& {\overline v}(p_{\overline e}) \gamma_\mu
 u(p_e) Q_l Q_q D_\gamma(P) {\overline u}(P/2,s_q) \gamma_\mu 
v(P/2,s_{\overline q})\nonumber\\
 & &+ {\overline v}(p_{\overline e}) \gamma_\mu (V_e + A_e \gamma_5)u(p_e) D_Z(P)
{\overline u}(P/2,s_q) \gamma_\mu (V_q + A_q\gamma_5)
v(P/2,s_{\overline q})
.
\ee
We take for the $Z$ propagator
\be
           D_Z(P) = \frac{1}{M_Z^2-P^2-iM_Z \Gamma_Z}
\ee
We use a constant $Z$ width equal to $2.5 \,GeV$ although our results are not
sensitive to this.  The photon propagator, $D_\gamma(P)$ is as above with 
mass and width put to zero.
Constant factors in the propagators do not affect our result.
The vector and axial vector couplings of the Z to a
fermion of weak isospin $I_{3f}$ and charge $Q_f$ are (relative to the
proton charge)
\be
     V_f &=&  (I_{3f}-2 Q_f \sin^2 \theta_W)/\sin(2 \theta_W) \nonumber\\
     A_f &=& - I_{3f}/\sin(2 \theta_W)
.
\label{couplings}
\ee     

   We define the quantization axis in the quarkonium rest frame to be the direction
of the incident lepton.  Covariantly, we put

\be
       s_{q \mu} = \lambda \frac{\delta_\mu}{M}\nonumber
\ee
\be
       s_{{\overline q} \mu} = {\overline \lambda} \frac{\delta_\mu}{M}
\ee
where
\be
       \delta_\mu = p_{e \mu} - p_{{\overline e}\mu}
\ee       
and $\lambda, \overline{\lambda}$ are $\pm 1$.
The polarization probabilities as a function of $\lambda, \overline{\lambda}$ are
then
\be
 P(\lambda,{\overline \lambda}) = \frac{b}{8 M^4} \sum |{\cal M}(\lambda,{\overline \lambda})|^2 
\ee
the sum being taken over the initial state lepton spins.  The proportionality constant
$b$ is fixed by requiring 
\be
P(\Uparrow)+P(\Downarrow)+P(0)=1
\label{norm}
\ee
 where
\be
    P(\Uparrow) &=& P(1,1)\nonumber \\
    P(\Downarrow) &=& P(-1,-1)\\
\nonumber
    P(0) &=& {\frac{1}{2}} (P(1,-1)+P(-1,1)
.
\ee
     Neglecting lepton masses we find that
\be
    P(\lambda,{\overline \lambda}) &=& \frac{b}{8 M^4} \,F_{\mu \nu} \left( M^2 g_{\mu \nu}(1+ \lambda {\overline 
 \lambda}) - i (\lambda+{\overline \lambda})\varepsilon(P,\delta,\mu,\nu) \right)
\ee
where
\be
 F_{\mu \nu} = -\frac{Q_q^2}{2} L_{\mu \nu}^\gamma |D_\gamma(P)|^2 
- \frac{V_q^2}{2}L_{\mu \nu}^Z 
|D_Z(P)|^2- 2 \frac{Q_q V_q}{2}L_{\mu \nu}^{\gamma Z}Re ( D_\gamma(P) D_Z(P)^*)
\ee

\be
     L_{\mu \nu}^\gamma = Q_e^2 Tr \left( \not\!p_{\overline e} \gamma_\mu 
\not\!p_e\gamma_\nu \right)
\ee
\be
L_{\mu \nu}^Z = (V_e^2+A_e^2) Tr \left( \not\!p_{\overline e} \gamma_\mu 
\not\!p_e \gamma_\nu \right) -4iV_e A_e \varepsilon(P,\delta,\mu,\nu)
\ee
\be
L_{\mu \nu}^{Z \gamma} = Q_e V_e Tr \left( \not\!p_{\overline e} \gamma_\mu 
\not\!p_e \gamma_\nu \right) -2iQ_e A_e \varepsilon(P,\delta,\mu,\nu)  .
\ee
Thus
\be
     P(\Uparrow)+P(\Downarrow) = b \left( Q_e^2 Q_q^2 |D_\gamma(P)|^2
+ V_q^2 (V_e^2+A_e^2) |D_Z(P)|^2 \right.\nonumber \\
+ \left. 2 Q_e Q_q V_e V_q Re(D_\gamma(P) D_Z^*(P)) \right)
\ee
\be
      P(\Uparrow)-P(\Downarrow) = 2 b V_q A_e \left(V_q V_e |D_Z(P)|^2 
+Q_q Q_e Re(D_\gamma(P) D_Z^*(P))\right)
\ee
\be
      P(0) = 0
.
\ee
The normalization condition eq.\,\ref{norm} implies that
\be
     b= \left( Q_e^2 Q_q^2 |D_\gamma(P)|^2 + V_q^2 (V_e^2+A_e^2) |D_Z(P)|^2 
+ 2 Q_e Q_q V_e V_q Re(D_\gamma(P) D_Z^*(P))\right)^{-1}
.
\ee

\section{\bf Forward Backward Asymmetry in Leptonic Quarkonium Decay}
\setcounter{equation}{0}

      The quarkonium polarization calculated in the previous section can be
tested in various decay final states.  For example, using the wave function factor of
eq.\,\ref{wffactor}, the amplitude for quarkonium decay into a lepton pair is
\be
  {\cal M} = 2 \pi \alpha \Psi(0) \sqrt{3 M} {\overline u}(p_f)\not \! \epsilon
   (V + A \gamma_5) v(p_{\overline f})
\ee      
where
\be
    V = Q_q Q_f D_\gamma(P) + V_q V_f D_Z(P) \nonumber \\
    A = V_q A_f D_Z(P)
\ee
the $V_{q,f}$ and  $A_{q,f}$ being given by eq.\,\ref{couplings}.
The matrix element squared is
\be
 |{\cal M}|^2 = \epsilon_\alpha(\lambda) \epsilon_\beta^*(\lambda)
    192 \pi^2 |\Psi(0)|^2 \alpha^2 M \left( (|V|^2+|A|^2) Tr( \not\!p_f
    \gamma_\alpha \not\!p_{\overline f} \gamma_\beta) \right.\nonumber \\
    +\left. (2 Re VA^*)Tr(\gamma_5 \not\!p_f \gamma_\alpha \not\!p_{\overline f}
    \gamma_\beta) \right)
.
\label{mesq}
\ee    
The probability weighted sum over quarkonium helicities is
\be
  {\overline {\sum_\lambda}} \epsilon_\alpha(\lambda) \epsilon^*_\beta(\lambda) =
 \frac{1}{2}(P(\Uparrow)+P(\Downarrow))\left(-g_{\alpha \beta}+ \frac{
P_\alpha P_\beta - \delta_\alpha \delta_\beta}{M^2} \right) 
&+& \frac{1}{2}(P(\Uparrow)-P(\Downarrow))\frac{-i \epsilon(P,\delta,\alpha
,\beta)}{M^2} \nonumber \\
+ P(0)\frac{\delta_\alpha \delta_\beta}{M^2}
.
\ee
This can be easily seen in the quarkonium rest frame using 
eqs.\,\ref{Up}-\ref{Zero}
and then generalized to an arbitrary frame using Lorentz covariance.
We write the final state fermion momenta in eq. \, \ref{mesq} as
\be
     p_f = (P+\delta_f)/2 \nonumber\\
     p_{\overline f} = (P-\delta_f)/2 .
\ee

Then, performing the quarkonium spin average in eq.\,\ref{mesq} yields
\be
  |{\overline{\cal M}}|^2 = 192 \pi^2 |\Psi(0)|^2 \alpha^2 M \left( 
(P(\Uparrow)+P(\Downarrow))(|V|^2+|A|^2)(M^2 + \frac{(\delta \cdot \delta_f) ^2}{M^2})
  \right. \nonumber\\ \left.
 -4 (P(\Uparrow)-P(\Downarrow)) Re(VA^*) \delta \cdot \delta_f
  + 2 P(0)(|V|^2+|A|^2) (M^2 - \frac{(\delta \cdot \delta_f) ^2}{M^2} ) \right) .
\ee
Here we have kept the $P(0)$ term for generality although it vanishes in quarkonium
production by lepton-antilepton annihilation neglecting the lepton mass.  
In terms of the angle $\theta$
between
the final state lepton and the initial state lepton in the quarkonium rest frame
\be
     \delta \cdot \delta_f = -4 \vec{p_e} \cdot \vec{p_f} = - M^2 \cos \theta
\ee
and
\be
  |{\overline{\cal M}}|^2 = 192 \pi^2 |\Psi(0)|^2 \alpha^2 M^3 \left( 
(P(\Uparrow)+P(\Downarrow))(|V|^2+|A|^2)(1+ \cos^2 \theta) \right. \nonumber \\ \left.
 +4 (P(\Uparrow)-P(\Downarrow)) Re(VA^*) \cos \theta
  + 2 P(0) (|V|^2+|A|^2) \sin^2 \theta \right) .
\ee
Putting $P(0)$ to zero, the forward backward asymmetry is
\be
     \frac{F-B}{F+B} &=& \frac{ \int_{-1}^1 d\cos\theta \,\epsilon(\cos\theta)\,|{\overline{\cal M}}|^2 }{ \int_{-1}^1 d\cos\theta \,|{\overline{\cal M}}|^2 }
 = \frac{3}{2}\frac{ Re VA^*}{|V|^2+|A|^2} (P(\Uparrow)-P(\Downarrow)) .
\label{FBasym}
\ee
     In table $1$ we collect at each of four quarkonia, the differences
$\Delta P = P(\Uparrow)-P(\Downarrow)$ and the 
forward-backward asymmetries from eq.\,\ref{FBasym}.  
For the vector toponium resonance we assume a mass of $340 GeV$.
The forward-backward asymmetries in the leptonic decay of the charm and bottom 
quarkonia might be measurable at future $J/\Psi$ 
or Upsilon facilities while the large predicted forward backward asymmetry
at toponium should be observable at the next linear collider.  The experiments, however,
are made difficult by the small branching ratios into lepton pairs.

\begin{table}
\centering
\be
\nonumber
\begin{array}{||c|c|c|c|c||}\hline
           & BR_\mu                   & \Delta P           &  (F-B) asym        &\, (F-B)asym_{off-res}\, \\\hline

    \Phi   &\quad (3.7 \pm 0.5)\cdot 10^{-4}\quad &\quad 1.82 \cdot 10^{-4} \quad &\quad 2.48 \cdot 10^{-8}\quad &-6.58 \cdot 10^{-5}\\ \hline

  J/\Psi   &.059 \pm .001             & 4.65 \cdot 10^{-4}&1.62 \cdot 10^{-7}&  -6.08 \cdot 10^{-4} \\\hline

 Upsilon   &.0248 \pm .0006           &1.59 \cdot 10^{-2}  &1.88 \cdot 10^{-4}&  -5.73  \cdot 10^{-3} \\\hline

 \quad Toponium \quad &                          &-0.409             & 0.125 & 0.496 \\ \hline   
\end{array}
\ee
\caption{Polarization differences and forward-backward asymmetries of lepton pair
decays of quarkonia.  Also tabulated are the experimental muon pair branching ratios and the 
theoretical off-resonance
forward-backward asymmetries.}
\end{table}

     Off-resonance, or neglecting the quarkonia effects, the forward backward asymmetries
are
\be
    \left(\frac{F-B}{F+B}\right)_{off-res} &=& \frac{3H}{4G}
\ee
where
\be
      G &=& |Q_e Q_f D_\gamma(P)|^2 + (V_e^2+A_e^2)(V_f^2+A_f)^2 |D_Z(P)|^2
          +2 Q_e Q_f V_e V_f Re(D_\gamma(P)D_Z^*(P)) \nonumber \\
      H &=& 2 A_e A_f \left( 2 V_e V_f |D_Z(P)|^2 + Q_e Q_f Re(D_\gamma(P) D_Z^*(P)) \right)  .
\ee
It is predicted that the forward backward asymmetry is opposite in sign and suppressed
in magnitude relative to the off-resonance values in the $\Phi, J/\Psi$, and $\Upsilon$
regions while in the toponium region the asymmetry retains its sign and is only slightly
suppressed in magnitude relative to off-resonance values.

{\em acknowledgments}

     This work was supported in part by the US Department of Energy
under grant number DE-FG02-96ER-40967.



\end{document}